\documentclass[12pt]{article}
\usepackage[hmargin=2.8cm,vmargin=3.3cm]{geometry}
  \expandafter\let\csname equation*\endcsname\relax
  \expandafter\let\csname endequation*\endcsname\relax
 \usepackage{graphicx}
 \usepackage{amsmath}
 \usepackage{amssymb}
  \usepackage{enumerate}
\usepackage{cite}
\usepackage{tikz}
\usetikzlibrary{decorations.pathmorphing}

\usepackage{color}

\usepackage{enumitem}
\usepackage{fix-cm}
\usepackage{array}
\usepackage{makeidx}
\usepackage{fix-cm}
\usepackage{cancel}
\usepackage{bigints}
\usepackage{relsize}
\usepackage{amsfonts}
\usepackage{latexsym}
\usepackage{amsthm}
\usepackage[all]{xy}
\usepackage{amscd}
\usepackage{marvosym}
\usepackage{float}
\usepackage{sidecap}
\usepackage{caption}
\usepackage{subcaption}
\usepackage{extpfeil}
\usepackage{chemarrow}
\usepackage{array}
\usepackage{setspace}
\newcommand{\be}{\begin{eqnarray}}
\newcommand{\ee}{\end{eqnarray}}

\begin{document}

 \title{Quantum statistical mechanics   near a black hole horizon}

 \author {  Eirini Sourtzinou\footnote{sourginou@upatras.gr} \; and Charis Anastopoulos\footnote{anastop@upatras.gr}\\
Department of Physics, University of Patras, 26504 Patras, Greece.}

\maketitle

\begin{abstract}
We undertake a first-principles analysis of the thermodynamics of a small body near a black hole horizon. In particular, we study the paradigmatic system of a quantum ideal gas in a small box  hovering over the Schwarzschild horizon. We  describe the gas in terms of  free  quantum fields, bosonic and fermionic,  massive and massless. We identify thermodynamic properties through the microcanonical distribution.
We first analyse the more general case of a box in Rindler spacetime, and then specialize to the black hole case.
The physics depends strongly on the distance of the box from the horizon, which we treat as a {\em macroscopic} thermodynamic variable. We find that the effective dimension of the system transitions from three-dimensional  to two-dimensional as we approach the horizon, that  Bekenstein's bound fails when the box is adiabatically lowered towards the black hole,  and that the pressure is highly anisotropic.  The pressure difference between the upper and lower wall leads to an effective force that must be added to the gravitational acceleration. We also show that the approximation of  quantum fields propagating  on a fixed background for matter breaks down when the system is brought to microscopic distances from the horizon, in which case backreaction effects must be included.
\end{abstract}

\section{Introduction}

\subsection{Motivation}
Bekenstein's proposal of black hole entropy \cite{Bek1}  and Hawking's derivation of black hole radiation \cite{Hawk}  signal our understanding of black holes as thermodynamic objects. Furthermore, the generalized second law of thermodynamics (GSL)\cite{Bek1, Bek2} asserts that black hole entropy adds up with matter entropy, and that their sum never decreases with time.

The GSL originates from the thermodynamic analysis of {\em small} bodies falling into a black hole, where by small we mean that their self-gravity is negligible.
This analysis assumes that the entropy of those small bodies is the same that they would have in Minkowski spacetime.
 This assumption appears reasonable if the dimensions of the body are much smaller than the curvature radius of the spacetime. Nonetheless, it is problematic;  black holes affect  the spacetime {\em causal structure}, and hence, the Hamiltonian of any small body in its vicinity.  Consequently, the presence of the horizon affects the density of states that defines the entropy in the microcanonical distribution.

In this paper, we undertake a first-principles analysis of the thermodynamics of a small body near a black hole horizon. The body under consideration is an ideal quantum gas  that is contained in a box  hovering above the horizon. The box is assumed to be sufficiently small, so that both self-gravity and the backreaction to the black hole is negligible. Then, the relevant theory for the particles in the box is
 Quantum Field Theory in curved spacetime (QFTCS). For non-interacting particles, the evolution equation of the fields are linear; hence, all properties of the QFTCS are determined by the structure of the field modes.
This allows us to evaluate the eigenvalues of the quantum field Hamiltonian, and from them to identify all thermodynamic properties of the system through the microcanonical distribution.  We find a strong dependence of thermodynamic variables on the distance of the box from the horizon.

This work is firstly motivated by the need to further understand the GSL, by  constructing more accurate models. Bekenstein has argued that the GSL is guaranteed if all thermodynamic systems satisfy an entropy bound (EB) \cite{Bek3}, namely, the ratio of the entropy $S$ to the energy $U$ obeys
\be
\frac{S}{U} \leq 2 \pi D. \label{BB}
\ee
  Here, $D$ is a characteristic  length scale, usually taken equal to the radius of a sphere that encloses the system.  There is independent evidence that this bound is satisfied in flat-space systems \cite{Bek4}, in weakly gravitating systems \cite{Bek5, Bousso}, and in strongly gravitating, static systems \cite{AnSav14}. However, there is as yet no analysis  for matter in the vicinity of a black hole horizon, which was the original set-up for Bekenstein's EB. We undertake such an analysis in this work.

Our second aim is to understand how  black hole horizons affect the statistical properties of matter. Quantum effects near the horizon may be manifested at a macroscopic scale, for example, as the quantum buoyancy force of Unruh and Wald \cite{UnWa1,  Bek8} that originates from Unruh radiation. This topic is significantly under-represented in the literature. We believe that it is important as a stepping stone towards a more nuanced description of the horizon that will also involve the effects of quantum backreaction.

Our third motivation comes from the foundations of statistical mechanics. It is far from obvious how to extend the usual recipes of statistical mechanics to general relativity, where there is no preferred notion of time translation \cite{WaldBH}. The fact that gravity is a long-range force complicates the thermodynamic description \cite{DRAW}, for example,  the canonical and the micro-canonical ensembles may be inequivalent.
In stationary spacetimes, a preferred notion of time translation exists, so it is possible to define a time-independent Hamiltonian, and hence, to analyze thermodynamic properties using
 the microcanonical distribution. Whether the standard  procedures  of statistical mechanics remain physically meaningful in  setups  so much removed from the theory's ordinary applications  is an open issue, that can be explored by the systems that we analyse in this paper.

\subsection{Relation to past work}

Padmanabhan et al   analyzed the thermodynamical properties of a {\em classical gas} in a  box near the horizon  \cite{Pad1, Pad2}. They used both a hydrodynamics description and a description in terms of classical statistical mechanics. These works look for an interplay between thermodynamic entropy of matter and
 the entropy of the horizon. For this reason they  focus on a  box that lies at a proper distance of the order of the Planck length from the horizon.
 Our analysis here considers {\em quantum} gases and covers a broader range of distances. We also show that backreaction cannot be ignored at small distances from the horizon, so that the idea of a test body at a Planck-scale proper distance  from the horizon is not consistent. In a related context, Ref. \cite{martinez} studied the thermodynamics of a classical box of gas in Rindler spacetime, but did not consider  the black hole regime.

In Ref. \cite{thooft}, 't Hooft  analyzed the statistical mechanics of a gas of massive particles inside a large spherical shell that surrounds a black hole. The internal radius of the shell is barely larger than the Schwarzschild radius, while the external radius is very large. As in our analysis, matter inside the shell is treated at the level of QFTCS, i.e., backreaction is ignored.
't Hooft identifies a divergent entropy term $S_{div}$ as the internal radius of the shell approaches the Schwarzschild radius, and  he eventually identifies $S_{div}$  with the Bekenstein-Hawking entropy of the black hole. The small box analysed here is very different from the shell in 't Hooft's model, in particular, it can be treated as a test body, for which backreaction is negligible. Still, the  spherical symmetry of the system allows us to reproduce $S_{div}$ from the entropy of the small box.

Our analysis is also influenced by models of equilibrium black holes, i.e., models of black holes inside a box that coexist with their Hawking radiation \cite{Davies, Hawk76, Page2, York, AnSav16}.
 Ref. \cite{AnSav16} incorporated backreaction in this setting, and showed that a thermodynamically consistent description of black holes in a box implies the existence of a thin but macroscopic shell around the horizon, where Einstein's equations fail. The shell likely consists of radiation with exotic thermodynamics---see, also Ref. \cite{LiLi92} for a similar conclusion in a different context. However,
 \cite{AnSav16} works only at the level of thermodynamics, and a deeper analysis, at the level of quantum statistical mechanics is required.

\subsection{Our results}
We study the thermodynamics of a quantum  gas in a box hovering above a Schwarzschild black hole horizon {\em at a proper distance significantly smaller than the Schwarzschild radius}. There are different levels of description for this system, in increasing degree of generality: (i) a purely hydrodynamic / macroscopic analysis, (ii) a microscopic analysis using classical physics, (iii) a microscopic analysis with quantum physics and (iv) a microscopic quantum description that incorporates backreaction and self-gravity effects. In this paper, we work at level (iii),   moving beyond past analyses that worked at  levels (i) and (ii). To this end, it is necessary to describe matter with QFTCS, which provides the only  consistent quantum description of matter in a background curved spacetime.
For free fields in a static spacetime, the QFTCS description is constructed solely from the properties of the single-particle Hamiltonian.
 We
 identify a general exact form for the density of states of this Hamiltonian, and then, we evaluate the relevant coefficients using a semi-classical approximation.
 We derive all thermodynamic quantities using standard methods from statistical mechanics.

The spacetime metric near the Schwarzschild horizon is approximately
Rindler. For this reason, we start our analysis from the thermodynamics of a box with matter in Rindler spacetime. This allows us to incorporate in our study the effects of acceleration on thermodynamics in settings other than black holes, for example, finding correction to the equations of state in presence of weak gravity.

\medskip

Our most important findings are the following.

\begin{enumerate}[leftmargin=*]
\itemsep0em
\item We construct the thermodynamic potentials for quantum fields in an accelerated cavity, modeled by imposing Dirichlet boundary conditions on free quantum fields in Rindler spacetime.   Thermodynamic quantities like entropy and internal energy are well defined.

\item We apply the results above to a small box near a Schwarzschild horizon. Then, we  check
 for the validity of the Bekenstein bound. In the simplest interpretation, where $D$ in Eq. (\ref{BB}) is identified with the proper radius of the box, we identify a regime of physically admissible parameters in which the bound is violated. In fact, a box that is lowered adiabatically towards the black hole will at some point violate Bekenstein's bound. We discuss possible ways that the bound may be restored.

\item For radiation, the scaling of entropy transitions from three-dimensional to {\em effectively} two-dimensional as we approach the horizon. For particles of mass $m$, and temperatures $T << m$, the system always  scales as two-dimensional near the horizon.

\item   In an accelerated box,   the pressure at the bottom wall is always larger than the pressure at the top wall.
The pressure differential leads to an effective acceleration on the box that is induced by the horizon. This is a novel phenomenon, we can only compare it with the Unruh-Wald buoyancy, even if the physical origins are very different. However, the acceleration we derive here points towards the horizon, and it adds  to the gravitational acceleration.

 \item Our approximations fail when we bring the box to microscopic distances from the horizon. The whole system of the box and  of the mechanism that keeps it static cannot be viewed as a test system, and its backreaction to the black hole must be taken into account. This means that the limit of taking the box to proper distance of order of the Planck length from the horizon is not well defined.

\end{enumerate}

The structure of this paper is the following. In Sec. 2, we analyse the quantum theory of a scalar field in a box in Rindler spacetime. Then, we consider the case of a Dirac field. In Sec. 3, we analyse the thermodynamic properties of an accelerated box.
In Sec. 4, we focus on gases of massless bosons  in the vicinity of the black hole horizon. In Sec. 5, we study massive bosons at low temperatures near the horizon. In Sec. 6, we summarize our results and discuss their implications.  In the Appendix, we extend our analysis to fermions.

\section{The quantum theory of a gas in a box in Rindler spacetime}

\subsection{Setup}
In this section, we construct the density of states of a bosonic field enclosed in a rectilinear box at rest  in Rindler spacetime with acceleration $a$
\be
ds^{2} \ = \ - a^{2} x^{2} \ dt^{2} \ + \ dx^{2} \ + \ dy^{2} \ + \ dz^{2}, \label{rindler}
\ee
expressed in terms of the standard coordinates $(t, x, y, z)$. The lowest wall of the box lies at $x_1$, and the upper wall at $x_2$. The proper length in the $x$ direction is $H_x = x_2 - x_1$.
 We denote the length of the box in the $y$ coordinate by $L_y$ and in the $z$ coordinate by $L_z$.

We analyse a
free scalar field inside the box, subject to the Klein-Gordon equation,
\be
\ddot{\Phi} - ax \ \partial_{x}(ax)\partial_{x}\Phi - (ax)^{2}\big{(}\partial_{x}^{2} + \partial_y^2 + \partial_z^2 - m^{2}\big{)}\Phi \ = \ 0 \ . \label{KG}
\ee
We solve the Klein-Gordon equation  subject to Dirichlet boundary conditions on  the box
\begin{eqnarray}
\Phi(x_1, z,t) = \Phi(x_2,y,z,t) \ = \ 0 \ , \nonumber \\
\Phi(x,0,z,t) = \Phi(x,L_{y},z,t) \ = \
\Phi(x,y,0,t) \ = \ \Phi(x,y,L_{z},t) \ = \ 0  \ .
\end{eqnarray}
The Dirichlet boundary conditions mimic walls that reflect particles  elastically.

 We look for mode solutions to Eq. (\ref{KG})  of the form $\exp [-i(\omega t - k_{y}y - k_{z}z)]  f(x)$. Substituting into Eq. (\ref{KG}), we obtain
\be
x^{2} f^{''} \ + \ x f^{'} \ + \ \bigg{[}\dfrac{\omega^{2}}{a^{2}} - \kappa^2 \ x^{2}\bigg{]} f \ = \ 0 \ , \label{fdiff}
\ee
where $\kappa = \sqrt{k_{y}^{2}+k_{z}^{2} + m^{2}}$.

The general solution of Eq. (\ref{fdiff}) is a linear combination of a modified Bessel function of the first type $I_{i\omega/a}(\kappa x)$ and of a modified Bessel function of the second type $K_{i\omega/a}(\kappa x)$ \cite{Unruhreview}.
The implementation of the Dirichlet boundary conditions with the modified Bessel functions leads to an algebraic equation from which we can determine the energy eigenvalues only numerically.  In this paper, we will obtain analytic expressions through the use of  a semi-classical approximation.

\subsection{The general form of the density of states}

The thermodynamic analysis of a  free-particle system does not require an explicit form of the energy eigenvalues, rather it suffices to identify the single-particle density of states at high energies.  We will show that the general form of this quantity can be determined without any approximation.
 We will only employ a semiclassical approximation in order to determine specific coefficients that appear in the density of states.  The semi-classical evaluation is accurate at high energies, hence gas temperatures. In Sec. 3, we will see that this approximation is reliable for temperatures significantly larger than the black hole's Hawking temperature.

We go back to Eq. (\ref{fdiff}), and we
 change variables to $\xi = \ln (ax)$. Eq. (\ref{fdiff}) becomes a Schr\"odinger-type equation
 \be
f''(\xi) + [E - U(\xi) ] f(\xi)= 0 \ , \label{WKB}
\ee
 with `energy' $E = \dfrac{\omega^{2}}{a^{2}}$ and `potential' $U(\xi) = \frac{ \kappa^2}{a^2} e^{2\xi}$.

 Let us denote by $V(\omega, \kappa)$  the number of eigenstates of Eq. (\ref{WKB})  with energy less than $\dfrac{\omega^{2}}{a^{2}}$.
 For large energies, where $E$ becomes approximately continuous, $V(\omega, \kappa)$  coincides with the area (actually length) of the classical energy surface
 \be
 \left(a \frac{k_{\xi}}{\omega}\right)^2 + \frac{ \kappa^2}{\omega^2} e^{2\xi} = 1 \nonumber
  \ee
  on the $\xi-k_\xi$ phase space. The symmetry of the energy surface under rescaling $k_{\xi} \rightarrow \lambda k_{\xi}$, $\kappa\rightarrow \lambda \kappa$, and $\omega \rightarrow \lambda \omega$, implies that the function $V(\omega, \kappa)$ is homogeneous of first order with respect to its arguments. Hence, we can express $V(\omega, \kappa)$ in the convenient form
 $V(\omega, \kappa) = \omega \zeta(\rho)$ for some function $\zeta$ of $\rho := (\kappa/\omega)^2$.

 Then, the number  $\Omega(\omega)$ of mode solutions to Eq. (\ref{KG}) with  energy less that $\omega$ is obtained by integrating $V(\omega, \kappa)$ with respect to $k_y$ and $k_z$,
 \begin{eqnarray}
 \Omega(\omega) \ = \ \frac{L_yL_z}{\pi^2}  \int dk_y dk_z V(\omega, \kappa) \ = \ \frac{L_yL_z}{\pi} \ \omega^3 \int_{\rho_{min}}^{\rho_{max}} d\rho \zeta(\rho) \ \ .
 \end{eqnarray}
 The maximal value $\rho_{max}$ is fixed by the boundary conditions. The minimal value $\rho_{min}$ is $m^2/\omega^2$ and it is obtained for  $k_y = k_z= 0$. Hence,
 \begin{eqnarray}
 \Omega(\omega) \ = \ \frac{L_yL_z}{\pi} \ \omega^3 \ [F(\rho_{max}) - F(m^2/\omega^2)] \ , \label{omeome}
 \end{eqnarray}
 \be
\text{where} \hspace{5cm} F(\rho) := \int_0^{\rho} d \rho' \zeta(\rho').\hspace{6cm} \label{fro}
   \ee
  The density of states $g(\omega)$ is the derivative of $\Omega(\omega)$.

  Eq. (\ref{omeome}) applies to the regime where $\omega$ can be treated as continuous, and it does not involve any approximation.  In particular, it implies that for
$m = 0$, $\Omega(\omega)$ is always proportional to $\omega^3$. In this case, it is convenient to define the effective length $\bar{L}_x$ in the $x$ direction as
 \vspace*{-\baselineskip}
\be
\bar{L}_x : =  \frac{3\pi}{2} F(\rho_{max}) \ , \label{effl}
\ee
so that
\be
\Omega(\omega) =  \frac{2\bar{L}_x L_yL_z}{3\pi^2} \omega^3 .
\ee
Then, the number-of-states function is formally identical with that of a gas of massless particles in flat spacetime: the Rindler acceleration has been fully incorporated in the definition of the effective length. The factor $\frac{2}{3}$ is due to the Dirichlet boundary conditions.

For Neumann or Robin boundary conditions on the box, the density of states $g(\omega)$ remains proportional to $  \omega^2$ and only a proportionality constant  changes. Thermodynamic variables   are  not significantly affected.

\subsection{Density of states in a semi-classical approximation}

Next, we evaluate the function $F(\rho)$ of Eq. (\ref{fro}) in a semi-classical approximation.
The Wigner-Kramers-Brillouin (WKB) solutions to Eq. (\ref{WKB}) are of the form
\begin{eqnarray}
f(\xi) \ = \ \left\{ \begin{array}{cc}  \dfrac{1}{\sqrt{k(\xi)}} \ \big{(}A e^{iS_{1}(\xi_1, \xi)} + B e^{-iS_{1}(\xi_1, \xi)}\big{)} \ , \ & \xi < \xi_{c} \\
  \dfrac{1}{\sqrt{\lambda(\xi)}} \ \Big{(}C e^{-S_{2}(\xi_c, \xi)} + D e^{S_{2}(\xi_c, \xi)}\Big{)} \ , \  &  \xi > \xi_{c} \end{array}\right. \label{fxi}
\end{eqnarray}
 where
 \be
k(s) &=& \sqrt{E-U(s)}\ ,  \hspace{0.6cm} \lambda(s) = \sqrt{U(s)-E} ,
 \ee
  \vspace*{-\baselineskip}
 \be
 S_{1}(\xi,\xi' )&=& \int_{\xi}^{\xi'}k(s)ds, \hspace{0.4cm} S_{2}(\xi,\xi' ) = \int_{\xi}^{\xi'}\lambda(s)ds ,
 \ee
 and   $\xi_c = - \ln \sqrt{\rho}$
 is the turning point. The form of the solution depends  on the relation of $\xi_c$ to $\xi_1$ and $\xi_2$. In particular, the box interior

\begin{itemize}
\item  is a classically forbidden region,  if $\xi_c < \xi_1$;
\item is a classically allowed region, if $\xi_c> \xi_2$;
\item contains a classically allowed and a classically forbidden region, if $\xi_1 < \xi_c < \xi_2$.
\end{itemize}

Since we focus  on the density of states $g(\omega)$ at the continuum limit, and not on a precise evaluation of all eigenvalues,  we can ignore the contribution from the classically forbidden regions. This approximation is often referred to as the {\em geometric optics approximation} to Schr\"odinger's equation.
Hence, we consider only the oscillating solutions in Eq. (\ref{fxi}) subject to the boundary conditions
\begin{itemize}
\item $f(\xi_1) =  0$
\item    $f(\xi_2) = 0$ if $\xi_c > \xi_2$;  $f(\xi_c) = 0$, if $\xi_c< \xi_2$, .
\end{itemize}

It follows that
  $\sin S_1(\xi_1, \xi_2) = 0$ for  $\xi_c > \xi_2$,  and
 $\sin S_1(\xi_1, \xi_c) = 0$  for  $\xi_c < \xi_2$.
Equivalently,   $S_1(\xi_1, \xi_2) = \ell \pi$ for $\xi_c > \xi_2$ and $S_1(\xi_1, \xi_c) = \ell \pi$ for $\xi_c <\xi_2$, $\ell = 1, 2, \ldots$.

Changing the integration variable in $S_1$ to $q:= \rho e^{2\xi}$, we obtain
 \begin{eqnarray}
 \frac{2 \ell \pi a}{\omega} =\int_{\rho s_1}^{c(\rho s_2)} dq \frac{\sqrt{1- q}}{q}, \label{eigenvo}
 \end{eqnarray}
where we defined $s_1 = (ax_1)^2  $, $s_2 = (ax_2)^2  $, and
\be
c(x) = \left\{ \begin{array}{cc} x \ , \ &  x < 1\\1 \ , \  &   \ \ \ x > 1 \ \ .
\end{array} \right.
\ee

\begin{figure}
 \includegraphics[height=4cm]{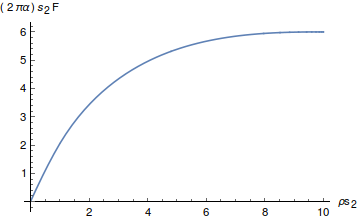}
    \caption{The function $F$ multiplied with $2 \pi a s_2$ as a function of $\rho s_2$ for of $s_2/s_1= 10$.}
    \label{fplot}
\end{figure}

Given  the eigenvalues (\ref{eigenvo}), we straightforwardly evaluate the number-of-states function  $V(\kappa, \omega) = \omega \zeta(\rho)$, where
\begin{eqnarray}
\zeta(\rho) &=&  \frac{1}{2 \pi a}  \int_{\rho s_1}^{c(\rho s_2)} dq \frac{\sqrt{1- q}}{q}. \  \
\end{eqnarray}
The maximum value of $\rho$ is
$\rho_{max} = \frac{1}{s_1}$.
Then,  the function $F(\rho)$ of Eq. (\ref{omeome}) is
\be
F(\rho) \ = \  \frac{1}{2 \pi a} \times \left\{ \begin{array}{cc} \frac{u(\rho s_2)}{s_2} - \frac{u(\rho s_1)}{s_1}, & \rho < \frac{1}{s_2}\\
\frac{u(1)}{s_2} - \frac{u(\rho s_1)}{s_1}  ,& \rho>  \frac{1}{s_2} \end{array} \right. \label{Fro}
\ee
where
\be
u(x) \ = \ \frac{2}{3} \sqrt{1-x}(1+2x)  -\frac{2}{3}  - 2 x \tanh^{-1}\sqrt{1-x} \  .
\ee
The function $F$ is increasing by construction; it is plotted in Fig. \ref{fplot}.

 Eq. (\ref{Fro}) implies that
\be
F(\rho_{max})   = F(s_1^{-1}) = \  \dfrac{1}{2 \pi a} \ \frac{2}{3} \ \Big{(}s_1^{-1}-s_2^{-1}\Big{)}\ . \label{From}
\ee
It follows that
\be
\bar{L}_x = \frac{1}{2a^3}(x_1^{-2} - x_2^{-2}).
\ee

\section{Thermodynamic properties of the accelerated box}
In this section, we analyse the thermodynamical properties of the accelerated box. We  consider  massless and massive bosons, and we also examine the  weak-acceleration limit.

\subsection{Entropy}
By Eqs. (\ref{omeome}) and (\ref{From}), the single-particle density of states for $m = 0$ is
\begin{eqnarray}
 g(\omega) := \frac{d \Omega(\omega)}{d \omega} =  \frac{2\bar{L}_{x} L_{y} L_{z}}{\pi^{2}}   \omega^{2} = \frac{2\bar{V}}{\pi^{2}}  \omega^{2}, \label{gome}
\end{eqnarray}
where we wrote  $\bar{V} = \bar{L}_{x} L_{y} L_{z}$.

 The density of states is given by the standard expression for photons in flat spacetime, modulo the substitution of $V$ with $\bar{V}$. The partition function is standardly evaluated \cite{Huang},
  \begin{eqnarray}
  \log Z(\beta,\bar{V}) = - \int_0^{\infty} d \omega g(\omega) \log (1 - e^{- \beta \omega}) = \beta \int_0^{\infty} d \omega \frac{\Omega(\omega)}{e^{\beta \omega}-1} = \frac{2\pi^2}{45} \bar{V} \beta^{-3}.
  \end{eqnarray}
In the canonical distribution, we identify $- \beta^{-1}\log Z$ with the Helmholtz free energy, and thereby obtain the thermodynamic description of the system. However,  the appropriate ensemble for an isolated box with adiabatic walls is the microcanonical one. Since acceleration renders the system non-extensive, the canonical and the microcanonical ensemble are not guaranteed to be equivalent \cite{DRAW}.

The partition function $Z(\beta)$ is the Laplace transform of the density of states function $\Gamma(E)$, which defines the microcanonical  entropy $ S(E) = \log \Gamma(E)$. We evaluate the inverse Laplace transform of  $Z(\beta)$  in the first-order saddle-point approximation \cite{Huang}, to obtain
 \be
 \Gamma(E) = \left[8\pi (2\pi^2/15)^{-1/4}E^{5/4}\bar{V}^{-1/4}\right]^{-1/2} \exp\left(\frac{4}{3} (2\pi^2/15)^{1/4} E^{3/4} \bar{V}^{1/4} \right).
 \ee
The standard expression for the photon entropy follows,  modulo terms of order $\ln E/E$ and $\ln \bar{V}/\bar{V}$,
\be
S(E,\bar{V}) = \frac{4}{3} \left(\frac{2\pi^2}{15}\right)^{1/4} E^{3/4} \bar{V}^{1/4}. \label{sent}
\ee

 Identifying $E$ with the internal energy $U$, we obtain results fully consistent with the canonical distribution.
 The difference is that
  the temperature is now {\em defined} as $T := (\partial U/\partial S)_{V}$, and we need make no reference to a heat bath. Since  energy  is defined with respect to asymptotic time translations, and  entropy is a scalar, the temperature $T$ refers to observers at infinity. Assuming   Tolman's law, we can define a local temperature $T_{loc}(x) = T\sqrt{-g_{00}(x)} = T/(ax)$ associated to planes of constant Rindler coordinate $x$ inside the box.

The equations for entropy and internal energy are
\be
S = \frac{8 \pi^{2}}{45}  \bar{V}  T^{3}, \hspace{1cm} U = \frac{2\pi^2}{15} \bar{V}  T^{4}. \label{sei}
\ee
 What changes from  flat space   is   the proportionality constant $\sigma$ (the Stefan-Boltzmann constant) in the relation $U = \sigma V T^4$, \vspace*{-\baselineskip}
\be
\sigma =  \frac{2\pi^2}{15}  \frac{\bar{L}_x}{L_x}.
\ee
{\em Weak acceleration limit.} The usual spacetime metric for a weak homogeneous gravitational field
 \vspace*{-\baselineskip}
\be
ds^2 = -(1 + a \bar{x})^2 dt^2 + \frac{d\bar{x}^2}{(1+a \bar{x})^2} + dy^2 + dz^2
\ee
can be brought into the Rindler form (\ref{rindler}) with a change of coordinates $x = \frac{1}{a}+\bar{x} + O(a)$. Since $\bar{x}$ is a Cartesian coordinate, the proper length $H_x$ is identified with the coordinate length $L_x$. We  shift the coordinate $\bar{x}$, so that $x_1$ corresponds to $\bar{x} = 0$. Then, $x_1 = \frac{1}{a}$, $x_2 = \frac{1}{a} + L_x$, and
\be
\bar{L}_x = L_x + \frac{3}{2}L_x a + O[\left(L_xa\right)^2].
\ee
Hence, black-body energy to leading order in $L_xa$ is $U = U_0 ( 1 + \frac{3}{2}L_x a )$,
where $U_0$ is the energy in absence of gravity. The relative size of the  correction term for a box of length $L_x \sim 100$m in the gravitational field of the Earth is of the order of $10^{-14}$.

\subsection{Pressure}
The entropy (\ref{sent}) is a function on the thermodynamic state space $\Gamma$, i.e., a manifold spanned by the variables $U, L_y, L_z, x_1$ and $x_2$. The `lengths' $L_x, L_y, L_z, $ and $\epsilon$ are coordinates of a submanifold $\Gamma_{m}$ of $\Gamma$ for the mechanical degrees of freedom. Any variation of the internal energy with fixed entropy defines an one-form on $\Gamma_m$
\be
(dU)_{\Gamma_m} = - \left(\frac{\partial U}{\partial y^a}\right)_S dy^a,  \label{pressureform}
\ee
where we used $y^a$ as a shorthand for the coordinates $(L_x, L_y, L_z, \epsilon)$.

All variables of $\Gamma_m$  are  lengths that corresponds to changes in the location of a wall of the box. The derivatives in Eq. (\ref{pressureform}) lead to the definition of pressures. For this reason, we will refer to $(dU)_{\Gamma_m}$ as the {\em pressure one-form}.
Since the system is not homogeneous, it cannot be described by a single scalar for pressure. Each vector field $X^a$ on $\Gamma_m$ defines a different change of the walls, hence, each contraction  $(dU)_{\Gamma_m}(X)$  corresponds to a different type of pressure.

We proceed to define pressures with respect to the Rindler coordinate system.
For the horizontal pressure $P_h$, we choose either of the vector fields $ X_1 = \dfrac{1}{H_xL_z}\dfrac{\partial }{\partial L_y}$ or  $X_2 = \dfrac{1}{H_xL_y}\dfrac{\partial }{\partial L_z}$.
 Then,
\be
  P_h := - \frac{1}{H_xL_z} \left(\frac{\partial U}{\partial L_y}\right)_{S, L_z, x_1,x_2 }  = \frac{1}{3} \rho, \label{ph}
\ee
where
\be
\rho = \frac{U}{H_xL_yL_z} = \frac{2\pi^2}{15} T^4 \frac{\bar{L}_x}{H_x}
\ee
 is the average energy density in the box.

\begin{figure}
 \includegraphics[height=6cm]{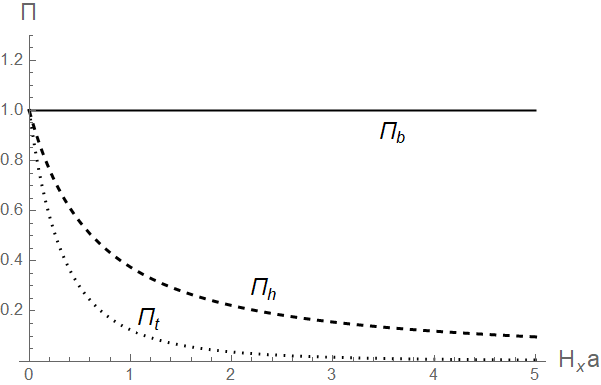}
    \caption{The quantities $\Pi_h, \Pi_b, \Pi_t$ expressed as functions of $H_xa$. The inequality (\ref{pressineq}) is evident.}
    \label{pressure}
\end{figure}

The pressure $P_b$ at the bottom of the box corresponds to a vector field $X_b =   \dfrac{1}{L_{y}L_{z}} \dfrac{\partial}{\partial x_1} $. Then,
 \vspace*{-\baselineskip}
\be
\hspace{-1.5cm} P_b :=   \frac{2\pi^2T^4}{15a^3x_1^3} = P_h\frac{x_2^2}{\frac{1}{2}(x_1+x_2)x_1}. \label{pb}
\ee
 Similarly, we define the pressure at the top of the box in terms of the vector field
  $X_t = -\dfrac{1}{L_{y}L_{z}} \dfrac{\partial}{\partial x_2} $,
 \vspace*{-\baselineskip}
\be
P_t :=  \frac{2\pi^2T^4}{15a^3x_2^3} = P_h \frac{x_1^2}{\frac{1}{2}(x_1+x_2)x_2}. \label{pt}
\ee
The three pressures always satisfy
\begin{eqnarray}
P_b > P_h > P_t. \label{pressineq}
\end{eqnarray}
We rescale coordinates so that $ax_1 =1$. Then, the quantities $\Pi_{h,b,t} = \frac{45a^3x_2^3}{2\pi^2T^4}P_{h,b.t}$ are functions solely of $aH_x$,
\be
\Pi_h = \frac{1 + \frac{1}{2} H_xa}{(1+H_xa)^2},\;\;\; \Pi_b = 1, \;\;\; \Pi_t = \frac{1}{(1+H_xa)^3}.
\ee
These quantities are plotted in Fig. \ref{pressure}.

\subsection{Gas of massive bosons}
For $m\neq 0 $, the number-of-states function has a more complex dependence on the energy $\omega$, and there are several different thermodynamic regimes depending on the relative values of the various parameters that characterize the system.  Here, we will consider the `non-relativistic' regime $T << m$.

For $m\neq 0$, Eq. (\ref{omeome}) implies that there is  minimum value of $\omega$,
\be
\omega_{min}(\epsilon) = m/\sqrt{\rho_{max}} = m \sqrt{s_1},
\ee
which functions as an effective rest mass of the bosons.

For $T << m$, the values of $\omega$ around $\omega_{min}$ dominate. We evaluate the number-of-states function (\ref{omeome}), by expanding  $F(\rho)$ around $\rho_{max}$. To this end, we define $\xi:= 1 - \rho/\rho_{max} \in [0, 1)$, and we expand
\be
 u(1-\xi) \simeq - \dfrac{2}{3} \ + \ \dfrac{4}{15} \ \xi^{5/2} \ + \ \dfrac{4}{35} \ \xi^{7/2} +  \mathcal{O}(\xi^{9/2}). {}\nonumber
\ee
Then, to leading order in $\xi$, we obtain
\be
F(s_1^{-1}) - F(\xi) = \left\{ \begin{array}{cc} \frac{2}{15\pi a s_1}\left[ \xi^{5/2} - \left(\frac{s_2}{s_1}\right)^{3/2} (\xi - 1 +\frac{s_1}{s_2})^{5/2}  \right],&   \xi > 1 - \frac{s_1}{s_2}\\
\frac{2}{15\pi a s_1}  \xi^{5/2}, &\xi \leq 1 - \frac{s_1}{s_2} \label{fsfs1}
 \end{array} \right.
\ee
It is convenient to express the number-of-states function $\Omega$ in terms of $q := \omega - \omega_{min}$, the effective kinetic energy of a particle. By definition, $ \dfrac{m^{2}}{\omega^{2}} \ = \ \rho_{max}(1- \xi)$, hence, $\omega/\omega_{min} = (1- \xi)^{-1/2} = 1 + q/\omega_{min}$.
Hence, to leading-order in $\xi$, \ $\xi \simeq 2q/(m\sqrt{s_{1}})$  \  and we write
 \vspace*{-\baselineskip}
\be
\Omega(\xi) = \frac{2L_yL_zm^3}{15\pi^2as_1} \left\{ \begin{array}{cc} \xi^{5/2} - \left(\frac{s_2}{s_1}\right)^{3/2} \left(\xi - 1 + \frac{s_1}{s_2}\right)^{5/2}& \xi > 1 - \frac{s_1}{s_2} \\
\xi^{5/2} & \xi< 1 - \frac{s_1}{s_2} \end{array} \right.. \label{fsfs2}
\ee
  The number-of-states function is very sensitive on the ratio $s_1/s_2$. The regime where $s_1/s_2 << 1$ is appropriate to a box near a black hole horizon, and it is explored in detail in Sec. 5.  Here, we consider the `classical' regime that corresponds to a dilute gas, in which case the partition function can be evaluated using the Boltzmann distribution.
  For a gas of $N$ particles, $Z = Z_1^N$, where $Z_1 = \int_0^{\infty}dq g(q) e^{-\beta (m\sqrt{s_1}+q)} = e^{-\beta m \sqrt{s_1}} \beta \int_0^{\infty} dq  \Omega(q) e^{-\beta q}$. We calculate,
\be
Z_1 = \sqrt{\frac{2}{\pi m}} \frac{L_yL_z}{\pi^2 \beta^{5/2} a (ax_1)^{3/4}} e^{-\beta m a x_1} \left[ 1 - \left(\frac{x_2}{x_1}\right)^3 e^{-\frac{1}{2}\beta m a x_1 (1 - x_1^2/x_2^2)} \right].
\ee
The partition function leads to analytic expressions for all thermodynamic observables. In particular, the
 internal energy per particle is
\be
U/N = max_1 + \frac{5}{2\beta} - \frac{\frac{1}{2}max_1(1- x_1^2/x_2^2)}{e^{\frac{1}{2}\beta m a x_1 (1 - x_1^2/x_2^2) }- 1}. \label{unr}
\ee
This quantity is an increasing function of $x_2/x_1$, and it converges to $ m ax_1 + \frac{5}{2 \beta}$ as $x_2/x_1\rightarrow \infty$.
In the weak gravity limit ($ax_1 = 1$, $x_2 = x_1 + L_x$),  $U/N = m + \frac{3}{2}T + \frac{1}{2} maL_x$,
i.e., the leading order correction is the dynamical energy of the center of the mass.

\begin{figure}
 \includegraphics[height=6cm]{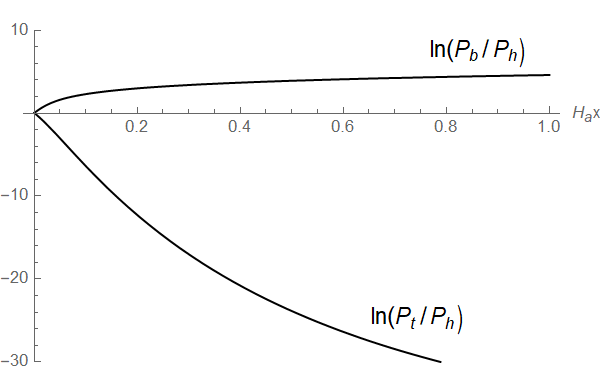}
    \caption{The logarithms of the ratios $P_t/P_h$ and $P_b/P_h$ as a function of $H_xa$ for $ax_1  = 1$. }
    \label{pressure2}
\end{figure}

We straightforwardly evaluate the pressures,
\be
P_h &=&\frac{N}{\beta H_x L_yL_z}, \\
P_b &=& P_h \, \beta m a H_x \left( 1 + \frac{\frac{3x_1^2}{2x_2^2} - \frac{1}{2}}{\left(\frac{x_1}{x_2}\right)^3e^{ \frac{1}{2}\beta m a x_1 (1 - x_1^2/x_2^2)}-1}\right),\\
P_t &=& P_h \,\beta m a H_x  \dfrac{\frac{x_1^3}{x_2^3} }{\left(\frac{x_1}{x_2}\right)^3e^{ \frac{1}{2}\beta m a x_1 (1 - x_1^2/x_2^2)} - 1}
\ee
In Fig. \ref{pressure2}, we plot the ratios $P_t/P_h$ and $P_b/P_h$. The pressures satisfy again inequality (\ref{pressineq}).

\section{Massless particles near a Schwarzschild horizon}
In this section, we apply the analysis of the previous section to a box  at rest near a black hole horizon, i.e., we assume that the box is accelerated so that the acceleration cancels free fall. We will work with the massless case here (radiation), and take up the massive case in the following section.

\subsection{Setup}
We consider a Schwarzschild spacetime of mass $M$,
\begin{eqnarray}
\hspace{-1cm} ds^{2}\ = \ - \Bigg{(}1-\dfrac{2GM}{r}\Bigg{)}dt^{2} \ + \ \Bigg{(}1-\dfrac{2GM}{r}\Bigg{)}^{-1} dr^{2} \ + \ r^{2}(d\theta^{2} + \sin^{2}\theta d\phi^{2}), \; \; \; \; \;\; \;
\end{eqnarray}
in terms of the standard coordinates $(t, r, \theta, \phi)$.

We assume that the box is much smaller than the Schwarzschild radius $r_{*}=2GM$. We take the lower wall  of the box to lie at radial coordinate $r = 2GM + \epsilon$, and the upper wall to lie at coordinate $r = 2GM + \epsilon + L_x$.

We assume that $\epsilon \ll 2GM$ and  that $L_x \ll 2GM$, but we make no assumption about the relation between $\epsilon$ and $L_x$. Then,
 the geometry near the horizon is approximately Rindler, as in Eq. (\ref{rindler}). In the present context,  $a = (4GM)^{-1}$ is the surface gravity and  $x = \sqrt{8GM(r- 2GM)}$ is the radial-proper length coordinate; $y$ and $z$ are local Cartesian coordinates on spheres of constant area.
  Hence, the proper length of the box is
\be
H_x =  \sqrt{8GM(\epsilon + L_{x})} - \sqrt{8GM\epsilon}.
\ee

The description of the local geometry near the horizon with the Rindler metric also applies to spherical symmetric dirty black holes \cite{Visser}, i.e., black holes that coexist with a spherically symmetric matter distribution. Hence, it also works for models of black holes in a box and in thermal equilibrium with their Hawking radiation as in Ref. \cite{AnSav16}.

The idealization of matter in a box does not make sense for black holes of stellar mass or smaller, because any physical box would be crashed by the strong tidal forces. However, such forces are much weaker in super-massive black holes, and in this case the idealization of a gas in a bounding box is physically meaningful. For a black hole with $M = 10^9 M_{\odot}$, the Schwarzschild radius is of the order of $10^9$km. Hence, the geometry in a box of proper length $1$m at proper distance $1$km from the horizon is well described by the metric (\ref{rindler}); and so is the geometry of a box of proper length $1$km at proper distance $1$m from the horizon.

We also note that the location of a bounding box in thermodynamics can only be viewed as a {\em macroscopic variable} (or a {\em macroscopic constraint} \cite{Callen}). Since a physical box is made of atoms, its location can only  be specified with an error larger than the atomic scale. Any suggestion that the error can be made as small as Planck's length has no physical justification. This is an important point because the models of Padmanabhan \cite{Pad1} and 't Hooft \cite{thooft} mainly concentrate on a box and a shell at Planck proper distance from the horizon. This is only possible if one invokes new physics of Planck-scale origin---'t Hooft explicitly makes such an assumption \cite{thooft2}.

In this section, we specialize to the case of massless particles and we assume that the particle number is not preserved. Then, our system is thermodynamically identical with  the photon gas, if we ignore the effects of polarization (scalar photons).

We will work in the thermodynamic state space $\Gamma$ that is spanned by the variables $U, L_y, L_z, L_x$ and $\epsilon$. To this end, we write
the effective length, Eq. (\ref{effl}), as a function of $\epsilon$ and $L_x$,
 \vspace*{-\baselineskip}
\be
  \bar{L}_x \ =  \ (2GM)^{2} \ \Bigg{(}\dfrac{1}{\epsilon} \ - \ \dfrac{1}{\epsilon + L_{x}}\Bigg{)} \  .
\ee
Substituting $\bar{L}_x$ into Eq. (\ref{sent}), we obtain the entropy as a function on $\Gamma$.

Note the importance of deriving Eq. (\ref{sent})   in the microcanonical distribution. The canonical distribution  presupposes that the walls of the box are diathermal, and that the box is surrounded by a thermal bath at temperature $\beta^{-1}$. The presence of such a bath would greatly complicate the analysis in a gravitational system, because it would gravitate and thus, change the spacetime geometry. A box in thermal contact with a gravitating thermal bath would not have the same thermodynamic properties with an isolated box in Schwarzschild spacetime \cite{KoAn21}. We also note that no thermal fluid can coexist with a black hole horizon in a spherically symmetric system \cite{AnSav21}. While there exist
 static, spherically symmetric solutions with Yang-Mills-Higgs and other types of fields outside the horizon \cite{nohv1, nohv2}, to the best of our knowledge, no such solution has an interpretation as a thermal bath.

 When quantum effects are included, the natural QFT vacuum  for a black hole in a box is the Hartle-Hawking vacuum. In this vacuum, Hawking radiation acts as a thermal bath of temperature
   $T_H = (8\pi GM)^{-1}$.
  However, the Hawking temperature is too low. We will show that the consistency of the statistical-mechanics analysis requires that $T >> T_H$. Note also that the Hawking temperature is fixed uniquely by the black hole mass, so in the canonical distribution we would be restricted to boxes of temperature $T_H$ in a given black hole. We conclude that the usual conditions for the validity of the canonical distributions are not available, and neither is the interpretation of   $\beta^{-1}$ as a temperature of a surrounding heat bath.

 It is convenient to express the pressures in terms of the Schwarzschild coordinates $\epsilon$ and $L_x$. Denoting the pressures in this frame by an overbar, we find that
  \be
 \bar{P}_h = \frac{H_x}{L_x} P_h,  \; \; \; \;  \bar{P}_b = \frac{x_1}{2\epsilon}P_b, \; \; \; \;
 \bar{P}_t = \frac{x_2} {2(L_x+\epsilon)}P_t. \label{rindpr}
 \ee

 \subsection{The small-box regime}
The  thermodynamics of the box with radiation depends crucially on the ratio $\epsilon / L_x$. First, we explore the regime where $\epsilon >> L_x$, i.e., we take the dimensions of the box to be  much smaller than its radial distance from the horizon.  Then, $\bar{L}_x \simeq \frac{8}{3}\left( \frac{GM}{\epsilon}\right)^{2} L_x$, and  $\bar{V} = \frac{8}{3}\left( \frac{GM}{\epsilon}\right)^{2} V$. Entropy and internal energy are proportional to the coordinate volume $V$. Hence, in this regime, energy and entropy behave like extensive quantities, and we can  define an energy density $\rho = U/ V$ and an entropy density $s = S/V$.
Furthermore,  $P_b \simeq P_h \simeq P_t$, i.e.,  pressure is almost isotropic, and the radiation satisfies the equation of state $P = \frac{1}{3}\rho$.

We cannot take the box to be arbitrarily small. In Minkowski spacetime,   the length $L$ of a box bounding radiation  must be much larger than the typical wave-length of the photons, which is of order $T^{-1}$. This is also necessary in order to describe the single-photon density of
states $g(\omega)$ as a continuous function. It follows that $TL >> 1$.

In the present context, the conditions $T L_y >> 1$ and $TL_z >>1$ follow  as in flat space. Regarding the $x$ direction, we note that the continuum approximation to the eigenvalues of Eq. (\ref{WKB}) follows from the requirement that $(\omega/a) (\xi_2 - \xi_1) >> 1$. The relevant scale for $\omega$ is given by $T$, hence,
\begin{eqnarray}
T >> \frac{1}{2GM\log(1+L_x/\epsilon)}. \label{TTH}
\end{eqnarray}
 For $L_x/\epsilon << 1$, Eq. (\ref{TTH}) implies that
\be
L_x T >> \frac{\epsilon}{2GM}. \label{Lxin}
\ee
Eq. (\ref{Lxin}) can also be written  as $L_x > > 4 \pi \epsilon (T_H/T)$, where
$T_H$ is the Hawking temperature. We conclude that a thermodynamic description for $L_x << \epsilon$ is meaningful only for temperatures  much larger than the black hole's Hawking temperature. The typical wavelength of Hawking radiation is of the order of the Schwarzschild radius, hence, it cannot be localized in a box of much smaller dimensions.


\subsection{The near-horizon regime}

If $\epsilon << L_x$, then $\bar{L}_x \sim \epsilon^{-1}$.  Internal energy and entropy diverge as $\epsilon \rightarrow 0$,
\be
S = \frac{8 \pi S_{BH}}{45} \frac{AT^3}{\epsilon},\hspace{1cm} U = \frac{2 \pi S_{BH}}{15} \frac{AT^4}{\epsilon}, \label{slimit}
\ee
where $S_{BH} = 4\pi G^2M^2$ is the Bekenstein-Hawking entropy of the black hole. However,  their ratio $S/U$ remains constant.

\begin{figure}
 \includegraphics[height=4cm]{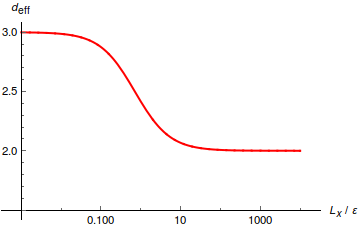}
    \caption{The effective scaling dimension of a box of radiation as a function of the ratio $L_x/\epsilon$.}
    \label{dimeff}
\end{figure}

For fixed $\epsilon$, both entropy and energy are proportional to the area $A = L_y L_z$. There is a transition from a   three-dimensional to a  two-dimensional system as we approach the horizon. We see this in Fig. \ref{dimeff}, where we plot the effective scaling dimension $d_{eff} := \log_2 \left[S(T, \epsilon, 2L_x, 2L_y, 2L_z)/S(T, \epsilon, L_x, L_y, L_z)\right]$ as a function of $L_x/\epsilon$.

However, we should refrain from interpreting the photon gas in this regime as a
  two-dimensional fluid, because  the pressure in the horizontal direction $P_h$ is much smaller than the bottom pressure $P_b$. The system is highly non-extensive: we would not be able to construct
  hydrodynamic variables (e.g., an effective stress tensor), even if we  restrict to a two dimensional brane around the horizon.

The divergence of the internal energy as $\epsilon \rightarrow 0$ implies that the approximation of treating the field as propagating on a background Schwarzschild spacetime breaks down near the horizon. Backreaction has to be taken into account. Keeping the box static near the horizon requires an external force. As the box approaches the horizon, the force becomes increasingly large and it must be incorporated into a stress-energy tensor which will deform the horizon.

We can ignore backreaction as long as the box can be treated as a test body in Schwarzschild spacetime, i.e., as long as $U << M$. By  Eq. (\ref{sei}), we obtain a constraint for the proper distance $x_1$ from the horizon,
\be
\frac{x_1}{l_{Pl}} >> (T \sqrt{A}) \left(\frac{M}{m_{Pl}}\right) \left(\frac{T}{T_{Pl}}\right), \label{rese}
\ee
  where $l_{Pl}, m_{Pl}$ and $T_{Pl}$, are the Planck length, Planck mass and Planck temperature respectively.

 For $T \sim 10^3K$, $M$ corresponding to a super-massive black hole and $L_y, L_z$ of the order of one centimeter, Eq. (\ref{rese}) implies that $\frac{x_1}{l_{Pl}} >> 10^{20} $, or, $x_1 >> 10^{-15} m$.
 From an operational point of view, the proper distance only takes macroscopic values, certainly it must be larger than the atomic scale, so the condition (\ref{rese}) poses no constraint. However, Eq. (\ref{rese}) demonstrates that we cannot take $x_1$ to be the Planck scale while ignoring backreaction.
We emphasize that inequality (\ref{rese}) refers solely to the validity of a QFTCS description without the inclusion of backreaction. The lower bound to $x_1$ does not reflect any intrinsic uncertainty in the location of the box, as in the corresponding  analyses of   Refs. \cite{Pad1} and \cite{thooft}.

Finally, we note that the consistency condition (\ref{TTH}) becomes in this regime
\be
T >> \frac{4 \pi T_H}{ \log(L_x/\epsilon)} = \frac{2 \pi T_H}{ \log(H_x/x_1)}. \label{TTH1}
\ee
The requirement that $T >> T_H$ suffices to guarantee the validity of Eq. (\ref{TTH1}). However, for sufficiently large values of $H_x/x_1$, a temperature $T \sim T_H$ might  be acceptable; the logarithmic dependence on $H_x/x_1$ makes it difficult to establish a sharp bound in an order-of-magnitude estimation.

The ratio $S/A$ given by Eq. (\ref{slimit}) coincides with the expression obtained by 't Hooft \cite{thooft} as the divergent term in the entropy of a large shell  around the horizon for $\epsilon \rightarrow 0$.  This coincidence is a consequence of spherical symmetry, and it occurs despite the fact that the two expressions are obtained for systems of different size and through calculations that involve sums over different modes.

Note that  the inner surface of the shell in  Ref. \cite{thooft} is taken at proper distance of the order of the Planck length from the horizon, while we found that in this regime the QFTCS description is inadequate and backreaction must be taken into account. Backreaction is a more severe problem in t' Hooft's model, as a shell can hardly be considered as a test body like the one considered here. The restriction that $U << M$ leads to a more stringent restriction for $x_1$ than Eq. (\ref{TTH1}) in 't Hooft's model for $T >> T_H$. However, it has been argued \cite{Israel} that for $T = T_H$, the inclusion of a negative contribution from the Boulware vacuum  renders the total energy sufficiently small, so that the QFTCS description suffices.




\subsection{Entropy bounds}

The Bekenstein bound  usually applies to {\em isolated}, weakly-gravitating systems, because of possible ambiguities in the definition of energy and of the characteristic length scale $D$ in other contexts. The system studied here is certainly not isolated, so the Bekenstein bound here, strictly speaking, does not apply here. Nonetheless, some entropy bound should exist also in the present context.
After all, entropy bounds originate from the requirement that the GSL applies also to setups that involve test bodies falling into the black hole.

Here, we proceed with the simple assumption that the length scale $D$ in Eq. (\ref{BB}) is the proper length of the smallest sphere that encloses the box, that is, $D = \frac{1}{2}\sqrt{H_x^2 + L_y^2 + L_z^2}$.
By Eq. (\ref{sei}), $\frac{S}{U} = \frac{4}{3T}$, hence,
 Eq. (\ref{BB}) becomes
\be
\frac{3 \pi}{4} T \sqrt{H_x^2 + L_y^2 + L_z^2} \geq 1. \label{BB2}
\ee
Clearly if $L_y, L_z$ are of order $H_x$ or larger, Eq. (\ref{BB2}) holds easily because $TL_y >> 1$  and $TL_z >> 1$. This is only possible if $L_x >> L_y, L_z$ and the box is not very close to the horizon.

This means that the contribution from the horizontal directions to any possible violation of the EB is negligible, so we can ignore it. Hence,  we look for possible violations of the inequality
 \vspace*{-\baselineskip}
\be
\frac{3 \pi}{4} T H_x \geq 1.  \label{BB3}
\ee
In the small box regime, Eq. (\ref{BB3}) becomes
\be
L_xT > \frac{4}{3\pi} \sqrt{\frac{\epsilon}{2GM}}. \label{BB4}
\ee
Eq. (\ref{BB4}) is compatible with the condition (\ref{Lxin}) for the thermodynamic limit, but it is not guaranteed by it.   Hence, there is a physically allowed regime of parameters, for which the EB in the form (\ref{BB2}) fails.

In the near-horizon regime, Eq. (\ref{BB3}) becomes $\frac{9 \pi}{16} (L_xT) (T/T_H) > 1$. The EB holds, if we can keep
 the temperature of the gas sufficiently large close to the horizon. This may not be always possible.


 \subsection{Adiabatic lowering of the box}
Suppose we prepare the box with the lower wall at some distance from the horizon. If the walls are adiabatic, then a quasi-static lowering of the box preserves the entropy $S$. By Eq. (\ref{sent}), the internal energy $U$ in an adiabatic process is proportional to $\bar{V}^{-1/3}$. This means that $U \sim \bar{L}_x^{-1/3}$. As the box approaches the horizon, $U \sim \epsilon^{1/3}$, i.e., the energy remains finite, and the approximation of the box as a test body remains accurate. This does not contradict Eq. (\ref{rese}), because the temperature $T$ also decreases when we lower the box adiabatically.

 The adiabatic lowering of the box leads to a decrease of the internal energy while entropy, and the proper length $H_x$ remain constant. Hence, if adiabatic lowering continues up to small distances from the horizon,   Bekenstein's bound will be violated. It is straightforward to show that in the near-horizon regime
\be
\frac{S}{U} = 2GM  \left(\frac{2^{11} \pi^2 A}{3^5 \ 5 \ S x_1^2}\right)^{1/3}.
\ee
In this regime, $D_x \simeq \frac{1}{2}\sqrt{8GML_x +L_y^2 + L_z^2}$ does not depend on $x_1$. Hence, the Bekenstein bound is violated for
\be
 x_1 <  1.3 \sqrt{\frac{A(GM)^3}{SH_x^{3}}}, \label{vbound}
\ee
which corresponds to macroscopic values of $x_1$.
 In fact, the Bekenstein bound fails well before we reach the near-horizon regime.

There are two possible explanations for this failure. The first explanation is that the condition (\ref{BB2}) is not an accurate representation of the entropy bound because the box is not isolated. We would have to go back to first principles, and find a new expression for the entropy bound that is valid for small material bodies near a black hole horizon. The other explanation is that the energy of the system must also include the energy of the walls enclosing it \cite{Bek6}. As we will see, the latter diverges as the system approaches the horizon, hence, its contribution would  keep the ratio $S/U$ small.

The average energy  density $\rho$ is proportional to $\bar{L}_x^{-1/3}$. As the box approaches the horizon, the horizontal pressure vanishes, $P_h \sim \epsilon^{1/3}$; the pressure at the top wall also vanishes, $P_t \sim \epsilon^{4/3}$; however, the pressure at the bottom diverges, $P_b \sim \epsilon^{-2/3}$.  Hence, the energy of the walls that keep the system confined increases as we approach the horizon. The Bekenstein bound (\ref{BB2}) may persist if the energy of the walls is included in the total energy\cite{Bek6}. To test this possibility, we need  an explicit model for the confining energy of the system---see, for example, \cite{Bousso2}.  However, we must point out that the divergence of the pressure and hence of the confinement energy occurs in the near-horizon regime, while the breakdown of Eq. (\ref{BB2}) occurs well before the system enters this regime.

In general, the fact that $P_b \neq P_t$ implies that there is a net total force acting on the box. We can evaluate this force by specifying how it is transmitted from the bottom to the top. Each mode of transmission corresponds to a vector field on $\Gamma_m$ that is contracted with the pressure form
 (\ref{pressureform}).  The simplest assumption is that the force is transmitted through {\em rigid} box walls, that is, the proper length $H_x$ remains constant in any displacement of the box.

Hence, for an adiabatic lowering of the box,  the acceleration due to the pressure differential is
 \vspace*{-\baselineskip}
\be
a_S := - \frac{1}{U} \left( \frac{\partial U}{\partial \epsilon}\right)_{S, H_x, L_y, L_z} =  - \frac{1}{3L_x} \left(1 + \frac{L_x}{\epsilon} - \frac{1}{\sqrt{1 + \frac{L_x}{\epsilon} }}\right) < 0.
\ee
The negative sign of $a_S$ means that the acceleration is directed towards the black hole.

Note that if the box is lowered isothermally, the acceleration is
\be
a_T := - \frac{1}{U}\left( \frac{\partial U}{\partial \epsilon}\right)_{T, H_x, L_y, L_z} = - 3 a_S,
\ee
i.e., it acts as  buoyancy, analogous to the one proposed by Unruh and Wald \cite{UnWa1}. However, as we mentioned, we are not aware of any static solution to Einstein's equation with a heat bath at temperature $T$ coexisting with  a black hole horizon. Hawking radiation could provide such a bath, but, as shown in Sec. 3.3.,  its temperature is so low that the usual thermodynamic approximations for the radiation in the box do not apply.

In the small-box regime, $a_S = - (6\epsilon)^{-1} $. This acceleration is comparable with the radial acceleration $a_r = - \epsilon^{-1}$ of a static  point-particle at radial distance $\epsilon$ from the horizon. Near the horizon, $a_S = -(3\epsilon)^{-1}$, and, expectedly, it diverges as $\epsilon \rightarrow 0$.

It is remarkable that, in the small-box regime, the acceleration $a_S$ does not depend on temperature or on properties of the box. It appears to depend on the properties of the particles--- for massive particles at low temperature, $a_S = - \epsilon^{-1}$---see below in Sec. 4.3. In this paper, we have seen this force only for static (or quasi-static) bodies. However, the mechanism that generates the force---the pressure difference in the radial direction---appears to be generic, and it is a consequence of the near-horizon geometry. Therefore, we conjecture that this acceleration could appear in other systems, for example, macroscopic bodies freely falling into the black hole. Since this acceleration depends on the properties of the gas, it is not equivalent to a change of geodesics due a quantum deformation of the metric, like, for example, that of Ref. \cite{fafa}. The derivation any observational consequences of this acceleration in astrophysical bodies would require a first-principles analysis, and an extension of our results to distances from the horizon larger than the ones considered in this paper.



\section{Massive particles near a Schwarzschild horizon}

\subsection{The number-of-states function }
We analyse  the thermodynamics of a box of massive particles (bosons and fermions) near the Schwarzschild horizon. We will mainly work in the regime where the temperature $T$ is much smaller than the mass $m$.
 This condition is also satisfied for Bose-Einstein condensates---see Ref. \cite{LWXY} for a study of Bose-Einstein condensates in the brick wall model.

Note that the opposite regime, of very high temperatures, $T >> m$, can be studied by expanding $F(m^2/\omega^2)$ in Eq. (\ref{omeome}) as a series in $(m/\omega)^2 << 1$. This leads to
\be
\Omega(\omega) = \frac{V}{\pi^2} \omega^3 + \frac{2GMAm^2}{\pi^2} \log(1 +L_x/\epsilon) \omega + \frac{4Vm^4}{3\pi^2} \omega^{-1} + \ldots,
\ee
where $A= L_y L_z$.

It is straightforward to show that the $m$-dependent terms only cause a small correction to the expressions for the thermodynamic variables in the massless case. There is no qualitative change, and for this reason, we will not elaborate on this regime.

To analyse  the regime of $T << m$, we first note that
the minimum value of $\omega$ for a box in the vicinity of the Schwarzschild horizon is
\be
\omega_{min}(\epsilon) =  m \sqrt{\frac{\epsilon}{2GM}},
\ee
i.e., $\omega_{min}$  vanishes for $\epsilon \rightarrow 0$.

By Eqs. (\ref{omeome}) and (\ref{fsfs1}), the number
$\Omega(q)$ of states with energy less than $\omega_{min} \ + \  q $  for a box close to the horizon is
\begin{eqnarray}
\Omega(q)  =  \eta(\epsilon)  A q^{5/2}.  \label{omegamm}
\end{eqnarray}
where
 \vspace*{-\baselineskip}
\begin{eqnarray}
 \eta(\epsilon) = \frac{64\sqrt{2}G^2M^2}{15\pi^2 \epsilon} \sqrt{\omega_{min}(\epsilon)} = \frac{16 \sqrt{2m}}{15 \pi^2}\dfrac{(2GM)^{7/4}}{\epsilon^{3/4}}. \label{condmassive}
\end{eqnarray}
This expansion of $\Omega$ applies as long as  $q << \omega_{min}$. Since the average value of $q$ is of the order of the temperature $T$, hence,  the expansion (\ref{omegamm}) is accurate for
\be
\frac{T}{m} << \sqrt{\frac{\epsilon}{2GM}}. \label{lalala}
\ee
This condition is much stronger than the usual condition $T << m$ for the non-relativistic regime. It implies that we cannot bring the box arbitrarily close to the horizon and still employ the expansion (\ref{omegamm}). The proper distance $x_1$ from the horizon satisfies $x_1/2 r_S >>  T/m $, where $r_S$ is the Schwarzschild radius.
For $T \sim 300 {}^oK$, particle mass $m \sim 100$ amu and $M \sim 10^9 M_{\odot}$, $x_1 >> 1$m.

It is straightforward to show that as the system approaches the horizon ($s_1 \rightarrow 0$), $F(m^2/\omega^2)$ approaches a constant $ - (3\pi s_2)^{-1}$ that is mass independent and much smaller than $F(\rho_{max})$;  $F(\rho_{max})$ diverges on the horizon. This means that the dominant contribution to the density of states near the horizon is of the form (\ref{gome}) that corresponds to massless particles. Hence, the near-horizon behavior is mass independent and coincides with that of Sec. 3.4.

 \subsection{Thermodynamic functions}

We proceed to the evaluation of thermodynamic observables using the grand-canonical distribution.
First, we write the number of particles
\begin{eqnarray}
\hspace{-1.5cm} N
=   \ \int_0^{\infty}  \frac{dq g(q)}{z^{-1}e^{\beta(\omega_{min} + q)}-1}  \ + N_0
= \dfrac{A}{\lambda_{2D}^{2}} \mbox{Li}_{5/2}(z e^{-\beta \omega_{min}(\epsilon)}) \ + \ N_0 \ ,\label{nmassive}
\end{eqnarray}
where $\mbox{Li}_a(x) = \sum_{n=1}^{\infty} \frac{x^n}{n^a}$ is the polylogarithm function, $N_0 = (z^{-1}e^{\beta \omega_{min}}-1)^{-1}$
is the average occupation number for the single-particle level with q = 0\footnote{This term originates from the divergence of the Bose-Einstein distribution at $ze^{-\beta \omega_{min}} = 1$, and it is evaluated prior to the continuous limit. Here, $\omega_{min}$ is defined as the lowest energy state after the continuous limit has been imposed. The actual minimum energy state $E_0$  differs from $\omega_{min}$ by a term of order unity. However, the explicit value of $E_0$ affects only the vacuum energy and pressure and not the other thermodynamic quantities of the system.}
, and
\be
\lambda_{2D}(\beta,\epsilon) = \sqrt{\frac{8 \beta^{5/2}} {15\sqrt{\pi} \eta(\epsilon)}}   \
\ee
is the effective de Broglie wave-length of the system. The label 2D emphasizes that in this regime the system becomes effectively two-dimensional, as the particle number scales with the area $A$ and not with the volume $V = L_x L_y L_z$, or with an effective volume like $\bar{V}$ that was defined for photons. Note that $\lambda_{2D} \sim \epsilon^{3/8}$, and it becomes increasingly smaller as we approach the horizon.

The grand potential
\begin{eqnarray}
\Phi  &=& \beta^{-1} \int_0^{\infty} dq g(q) \log(1- ze^{-\beta(\omega_{min} + q)}\ ) \ + \beta^{-1} \log(1 + N_0)  \nonumber\\
\nonumber\\
&=&  -\dfrac{A\beta^{-1}}{\lambda_{2D}^{2}}   \mbox{Li}_{7/2}(z e^{-\beta \omega_{min}(\epsilon)}) + \beta^{-1}  \ \log(1 + N_0), \label{gpf}
\end{eqnarray}
is also proportional to  the area $A$, modulo the ground state contribution $\log(1 + N_0)$.

The lowest energy for a single particle is $\omega_{min}(\epsilon)$, so the internal energy has a vacuum contribution $N \omega_{min}(\epsilon)$,
\begin{eqnarray}
U =     N \omega_{min}(\epsilon)  - \ \frac{5}{2 \beta} \Big{(}\Phi \ - \ \log(1 + N_0)\Big{)}. \; \;\; \; \; \; \label{mU}
\end{eqnarray}
The internal energy splits as $U = U_0 + U_{th}$, where $U_0 = N \omega_{min}(\epsilon)  $ is the ground-state energy and {\small{$U_{th} = -\dfrac{5}{2\beta} \Big{(}\Phi \ - \ \log(1 + N_0)\Big{)}$}} is the thermal energy; we call it thermal because it corresponds to the kinetic energy of the particles.

We also evaluate the entropy
\begin{eqnarray}
S &=& \dfrac{A}{(\lambda_{2D})^{2}} \left[(\beta \omega_{min}- \log z) \mbox{Li}_{5/2}(z e^{- \beta\omega_{min}})  + \frac{7}{2} \mbox{Li}_{7/2}(z e^{-\beta\omega_{min}}) \right]\nonumber\\
\nonumber\\[10pt]
&{}& + \left[ (N_0 + 1) \log (N_0 + 1) - N_0 \log N_0\right].
\end{eqnarray}
The height of the box $L_x$ does not appear explicitly in the equation for entropy. It drops out completely in this regime. This is the reason why the system scales with area and behaves essentially as two dimensional. When the temperature drops to the regime $T << m$, all gas molecules are constrained to move along the bottom of the box.

By Eq. (\ref{nmassive}), if  $N \lambda_{2D}^2/A \ > \ \zeta(\frac{5}{2})$, then $N_0 > 0$, i.e., there is a non-zero number of particles in the Bose-Einstein condensate phase. This condition can be written equivalently as   $T < T_c$, where
\be
T_{c} =  \frac{\pi^{3/5}}{2 [2\zeta(\frac{5}{2})]^{2/5}} \left(\frac{N}{A}\right)^{\frac{2}{5}} \left(\frac{2\epsilon^3}{(GM)^7 m^2}\right)^{\frac{1}{10}}.
\label{tc}
\ee
is the critical temperature for Bose-Einstein condensation. It follows that
\be
\dfrac{N_0}{N} = \begin{cases}
{} \ \ \ 0  \ , \ \   \ \  \ \    \ T \geq T_{c}\\
\\[2pt]
{} \ \ \ 1 \ - \ \Big{(}\dfrac{T}{T_{c}}\Big{)}^{5/2} \ , \ T \leq T_{c} \ \ . \label{mn0}
\end{cases}
\ee


\subsection{Specific regimes}

{\em Dilute gas. }
The dilute gas regime corresponds to $ z e^{-\beta\omega_{min}(\epsilon)} \ll 1$, which implies that  $\mbox{Li}_{\alpha}(z') \simeq z'$. There is no Bose-Einstein condensate. We obtain
\be
N \simeq \dfrac{A}{\lambda_{2D}^{2}} \ z e^{-\beta\omega_{min}(\epsilon)} \hspace{0.4cm} , \hspace{0.6cm} U_{th} \ \simeq   \dfrac{5}{2}NT \ .  \label{nu}
\ee
The entropy is
\be
\hspace{-1.5cm} S = \ \dfrac{7}{2}N  +  N \log\Bigg{(} \dfrac{A}{N \lambda_{2D}^{2}}\Bigg{)} =  \dfrac{7}{2}N \ + \ N\log\Bigg{[}\Bigg{(}\dfrac{15\sqrt{\pi}A \eta(\epsilon)}{8N}\Bigg{)}\Bigg{(}\dfrac{2}{5} \ \dfrac{U_{th}}{N}\Bigg{)}^{5/2} \Bigg{]} . \label{dgs}
\ee
Since  $L_{x}$ does not appear explicitly in the fundamental equation (\ref{dgs}), there is no pressure at the top of the box.
It is straightforward to calculate the  horizontal pressure
\be
  \bar{P}_h  = \frac{2 U_{th}}{5 V} = \frac{NT}{V}\ ,
\ee
and the pressure at the bottom
\be
\bar{P}_b  =  \dfrac{N \omega_{min}}{2 A \epsilon }  +  \dfrac{3U_{th}}{10A \epsilon} =  \dfrac{N }{2 A \epsilon } \left(\omega_{min} + \frac{3}{2}NT\right) \ . \label{pbd}
\ee
We also evaluate the acceleration on a box that is lowered adiabatically
\be
a_{S}  =  - \frac{L_x}{2\epsilon (L_x + \epsilon) \left( 1 - \frac{1}{\sqrt{1+\frac{L_x}{\epsilon}}}\right)} + O(T/m).
\ee
This expression differs  from that for massless or ultra-relativistic particles.
In the small-box regime, $a_S = -\epsilon^{-1}$ and in the near-horizon regime, $a_S = - (2\epsilon)^{-1}$.

\medskip

\noindent {\em Dense gas.}
Bose-Einstein condensates correspond to  $z = 1$.  In this case,  entropy and thermal energy scale with the number of particles in the gas phase,
\be
N - N_0 = \zeta(5/2)  \dfrac{A}{\lambda_{2D}^{2}}, \; \; \;
U_{th} =   \ \dfrac{5 \zeta(7/2)}{2 \zeta(5/2) } (N- N_0) T, \;\;\;
S =  \frac{7U_{th}}{5T }.
\ee
 The equations for pressure and acceleration are the same as in the dilute gas case.

The Bose-Einstein condensate phase exists for $T < T_c$. A thermodynamic description is possible only if $T_c >> T_H$, which implies that
\be
\epsilon >> \frac{[\zeta(5/2)^{4/3}]}{2^{17/3}\pi^2}\frac{m^{2/3}}{GM (N/A)^{4/3}} \simeq 0.003 \frac{m^{2/3}}{GM (N/A)^{4/3}} ,
\ee
or equivalently, $x_1 >> 0.5 \frac{m^{1/3}}{(N/A)^{2/3}}$.

For  $m = 100$amu and $\sqrt{A/N} \sim 3$ nm (air in room temperature), we find that   $x_1 >> 10^{-34}$m, i.e., no significant constraint to $x_1$.   Note that by Eq.  (\ref{lalala}),  the Bose-Einstein condensate phase cannot persist in the near-horizon regime.

\medskip

\noindent
{\em Entropy bounds.}
In the regime where $L_x >> \epsilon$, the  gas of massive particles and the gas of massless particles have the same thermodynamic behavior. Hence, as shown in Sec. 3.5, the EB is violated.

For a dilute gas and for $L_x >> \epsilon$, Eqs. (\ref{nu}) and (\ref{dgs}) apply. We note that entropy $S$ varies as  $\log \epsilon$  while $U \simeq N \omega_{min}$ decreases with $\sqrt{\epsilon}$. Hence, we expect that there is a range of parameters such that the EB is violated when we bring the box sufficiently close to the horizon, while preserving the condition $L_x >> \epsilon$. The reader can straightforwardly verify a violation of the EB for $T = 1K$, $H_x = L_y = L_z = 1$cm and $x_1 = 1$m.
 Again, the  EB is eventually violated when we lower a box adiabatically, where $U$ decreases with $\sqrt{\epsilon}$ while $S$ remains constant.

In contrast, the entropy content of a Bose-Einstein condensate is negligible, hence, there is no violation of the EB, as long as $L_x >> \epsilon$.

\section{Conclusions}
We presented our motivation and our results in the Introduction. Here, we comment on the implications of our work.

First, the combination of QFTCS for the ideal gas dynamics and of the microcanonical distribution leads to a self-consistent description of the systems that were studied here. We expect that the results can be generalized to larger class of small systems near the black hole horizon, including  self-gravitating systems. We saw that there is a large and yet unexplored thermodynamic phenomenology of such systems, including the presence of acceleration due to anisotropic pressure. Its detailed
study will enable, among others,  a deeper understanding of the GSL, especially its grounding on the quantum-field theoretic description of thermodynamic systems.

Second, our results highlight the importance of backreaction for any discussion of horizon  thermodynamics, as QFTCS fails for macroscopic distances beyond the horizon, well before we approach the Planck length. We expect that a combination of the methods of Ref. \cite{ AnSav16} and the methods developed here could prove fruitful for implementing backreaction.

Finally, we note that the formulation of entropy bounds for small systems near a black hole need to be reappraised. We saw that the simplest implementation of the Bekenstein bound fails, possibly because this bound applies primarily to isolated systems. It is important to understand how or whether this failure can be remedied,  before proceeding to the formulation of a general entropy bound applicable to this class of systems.

\begin{appendix}

\section{The case of fermions}
In this Appendix, we analyze the case of a box with a fermion gas near a black hole horizon. The overall behavior is very similar with the bosonic case analyzed in the main text, the differences are primarily technical.
\subsection{Density of states for fermions}
The Dirac equation in a curved spacetime with metric $g_{\mu\nu}$ reads
\be
[i\gamma^{\mu}(\partial_{\mu} - \Gamma_{\mu}) - m] \hat{\psi} = 0,
\ee
where the curved-spacetime Dirac matrices $\gamma^{\mu}$  are defined in terms of the Minkowski Dirac matrices $\gamma_a$ by  $\gamma^{\mu}= E^{\mu}_a\gamma^{a}$. Here,
 $E^{\mu}_a$ the inverse of the tetrad field associated to $g_{\mu \nu}$, and $\Gamma_{\mu} := \frac{1}{8} E^{\nu}_a E_{b\nu; \mu}[\gamma^a, \gamma^b]$ is the spin connection.

For the Rindler metric, the Dirac equation becomes
\be
a^{-1}  \frac{\partial}{\partial t} \hat{\psi} = ( - i x \; {\bf \alpha} \cdot {\bf \nabla}  - \frac{i}{2} a^1 \psi + m x \beta) \hat{\psi},
\ee
where $\alpha^i = \gamma^0 \gamma^i$ and $\beta = \gamma^0$.
We consider modes of the form $e^{-i \omega t + i k_y y + ik_z z} f(x)$, where $f(x)$ is a Dirac spinor. One can show \cite{SMG, Oritti} that there are two linearly independent positive-frequency solutions for $f$, and these are obtained from the solution of the scalar equation
 \vspace*{-\baselineskip}
\be
x^{2} f^{''} \ + \ x f^{'} \ + \ \bigg{[}\left(\dfrac{\omega}{a}\pm \frac{i}{2}\right)^2 - \kappa^2 \ x^{2}\bigg{]} f \ = \ 0 \ , \label{fdiff2}
\ee
The modes $f$ still satisfy  Schr\"odinger's equation (\ref{WKB}), albeit with complex energy $E = \left(\dfrac{\omega}{a}\pm \frac{i}{2}\right)^2 = \frac{\omega^2}{\alpha^2} - \frac{1}{4} \pm i \frac{\omega}{a} $. For  a complex energy eigenvalue $E_R + i E_I$ the WKB solutions are of the form $Re^{\pm iS}$, where now
\be
R= \frac{1}{4} \int dx' \frac{-V'(x') + E_I}{E_R - V(x')},
\ee
but the Hamilton-Jacobi action $S$  is unaffected. Therefore, the eigenvalue equation (\ref{eigenvo}) remains unchanged modulo the change of $\omega$ with $\sqrt{\omega^2 - \frac{1}{4}a^2}$, together with the restriction that $\omega > \frac{1}{2}a$.

This change in $\omega$ is negligible for any temperature  $T$ that are much larger than the Unruh temperature $T_U = \frac{a}{2\pi}$ of the accelerated box. As shown in Sec. 4.2, the condition $T >> T_U$ is also necessary for the validity of the semi-classical approximation and  of the thermodynamic description.   Hence, the density of states is simply twice that of scalar particles, corresponding to the two spin directions of the Dirac particle.

\subsection{Density of states}

We will first analyse the case of ultra-relativistic fermions ($m= 0$). The density of states is $g(\omega) = \frac{4\bar{V}}{\pi^{2}}  \omega^2$ (twice that of the scalar case).  We will use  the grand-canonical distribution.  For ideal gases, and in the limit of a large number of particles, the grand canonical distribution gives the same entropy with the microcanonical distribution \cite{Huang}. However, in the present context, the parameters $\beta$ and $z$ that appear in the grand-canonical distribution (see below)  are not to be viewed as the inverse temperature and the fugacity of an external heat bath, but as arbitrary constants (Lagrange multipliers).

 The grand potential $\Phi$ and the particle number $N$ for one particle species are defined as
 \vspace*{-\baselineskip}
\be
\Phi &=& - \beta^{-1} \int d \omega g(\omega) \ln(1 + ze^{-\beta \omega}) = -\int d \omega \frac{\Omega (\omega)}{z^{-1}e^{\beta \omega}+1} \\
N &=& \int d \omega \frac{g(\omega)}{z^{-1}e^{\beta \omega}+1},
\ee
where $z = e^{\beta \mu}$ with $\mu$ the chemical potential.

For ultra-relativistic fermions, the particle number is not preserved in equilibrium. What is preserved is the lepton number $N_l = N - \bar{N}$, where $N$ is the number of particles and $\bar{N}$ is the number of anti-particles. The grand potential is a sum of the grand potential for particles and for antiparticles, while the chemical potential of anti-particles is the opposite of the chemical potential for particles. Then, we calculate
\be
\Phi = -\frac{8 \bar{V}}{\pi^2 \beta^4}[f_4(z) + f_4(z^{-1})], \label{phifer} \\
N_l = \frac{8\bar{V}}{\pi^2 \beta^3}[f_3(z) - f_3(z^{-1})], \label{lfer}
\ee
We identify the
three pressures    $P_h := - \frac{1}{H_xL_z} \left(\frac{\partial \Phi}{\partial L_y}\right)_{z, \beta, L_z, x_1,x_2 }$,  $P_b := - \frac{1}{L_yL_z} \left(\frac{\partial \Phi}{\partial x_1}\right)_{z, \beta, L_y, L_z, x_2 }$ and  $P_t := - \frac{1}{L_yL_z} \left(\frac{\partial \Phi}{\partial x_2}\right)_{z, \beta, L_y, L_z, x_1}$. Since $\Phi$ depends on the length parameters only through $\bar{V}$, Eqs. (\ref{ph},\ref{pb}, \ref{pt})  still apply. The only difference from the bosonic case is the explicit expression for the energy density $\rho = U/(H_xL_yL_z)$, where $U = -3\Phi$,
\be
\rho = \frac{24T^4}{\pi^2} \frac{\bar{L}_x}{H_x} [f_4(z) + f_4(z^{-1})],
\ee
where $z$ is expressed in terms of $N_l$ by Eq. (\ref{lfer}).

We also consider massive fermions at zero temperature. To be precise, we consider finite   temperature $T >>T_U$, but  small enough so that the entropy of the gas is negligible. In this regime, there is no particle-antiparticle pair creation, hence, the number of particles $N$ is conserved. The chemical potential coincides with the Fermi energy $\epsilon_F$, defined by $\Omega(\epsilon_F) = N$. For $\Omega$ given by twice the expression (\ref{fsfs2}), we obtain
\be
\Phi = -\frac{4L_yL_zm^3}{105 \pi^2 a^2 x_1} \xi_F^{7/2} \times \left\{\begin{array}{cc} 1 -  \left(\frac{x_2}{x_1}\right)^3\left[ 1 - \frac{1- (x_1/x_2)^2}{\xi_F}\right]^{7/2}, & \xi_F  \geq 1 - \frac{x_1^2}{x_2^2}\\1&\xi_F < 1 - \frac{x_1^2}{x_2^2} \end{array} \right., \label{phi00}
\ee
where $\xi_F$ is defined by the condition $\Omega(\xi_F) = N$. Given Eq.(\ref{phi00}), it is straightforward to compute all thermodynamic quantities, including the pressures. The latter simplify significantly in the regime where $\xi_F < 1 - \frac{x_1^2}{x_2^2} $ that is, as we will see, relevant to the black hole case. We find that
\be
P_h = -\frac{\Phi}{H_xL_yL_z},\;\;\, P_b = -\frac{\Phi}{x_1L_yL_z},\;\;\; P_t = 0.
\ee

\subsection{Gas of massless fermions}

There are few differences in the behavior of a fermionic gas of massless particles near the horizon. As shown in Sec. 3.4, the pressures behave exactly as in the bosonic case. The only difference is the dependence of entropy and total energy on the total lepton number $N_l$. In the grand ensemble, $S = (U - \Phi - \mu N_l)/T$. Since $U = -3\Phi$, Eqs. (\ref{phifer},\ref{lfer}) yield
 \vspace*{-\baselineskip}
\be
U &=& \frac{24\bar{V}T^4}{\pi^2} [f_4(z) + f_4(z^{-1})] \\
S &=&  \frac{8\bar{V}T^3}{\pi^2}\left[4[f_4(z) + f_4(z^{-1})] - \ln z [f_3(z) - f_3(-z)]\right].
\ee
To express $U$ and $S$ as a function of $N_l$, we must solve Eq. (\ref{lfer}) for  $z$ and substitute in the equations above.

At  high-temperatures, $z \simeq 1 + \nu$, where $\nu << 1$, and $f_a(z) = f_a(1) + \nu f'_{a_1}(1)$. Then, the dominant contribution to $U$ and $S$ are $N_l$-independent
\be
U = \frac{7\pi^2\bar{V}T^4}{15}, \hspace{1cm} S =  \frac{28\pi^2\bar{V}T^4}{45}
\ee
Hence, the analysis of the breakdown of the Bekenstein bound and of the QFTCS approximation remain unchanged modulo numerical constants of order unity.

In the low temperature regime, either $z \rightarrow \infty$ or $z \rightarrow 0$. We will consider the case $z \rightarrow \infty$, which corresponds to a surplus of particles. The extension to $z \rightarrow 0$ (surplus of antiparticles) is straightforward. In this regime, $f_n(z) \simeq (\ln z)^n/n! + \frac{\pi^2}{6(n-2)!} (\ln z)^{n-2}$ and $f_n(z^{-1}) \simeq z^{-1}$. We straightforwardly solve Eq. (\ref{lfer}) for $z$,
\be
\ln z = \left(\frac{3\pi^2 N}{4\bar{V}T^3}\right)^{1/3} - \frac{\pi^2}{3}\left(\frac{3\pi^2 N}{4\bar{V}T^3}\right)^{-1/3}.
\ee
Then, we find
 \vspace*{-\baselineskip}
\be
U  &=& \frac{3}{4} (3\pi^2/4)^{1/3} \frac{N_l^{4/3}}{\bar{V}^{1/3}} + \frac{1}{6}(3\pi^2/4)^{2/3} T^2 N_l^{2/3} \bar{V}^{1/3} , \\
 S &=& \frac{2}{3} (3\pi^2/4)^{2/3} T N_l^{2/3} \bar{V}^{1/3}. \label{entgg}
\ee
In the near horizon regime,
\be
U &=& \frac{3^{5/3} \pi}{8} \frac{N_l^{4/3} }{A^{1/3}S_{BH}^{1/3}} \epsilon^{1/3} + \frac{3^{1/3} \pi}{12} T^2N_l^{2/3}S_{BH}^{1/3} A^{1/3} \epsilon^{-1/3} ,
\\
S &=&\frac{3^{1/3} \pi}{3} TN_l^{2/3}S_{BH}^{1/3} A^{1/3} \epsilon^{-1/3} .
\ee
The dominant contribution to the internal energy vanishes as $\epsilon \rightarrow 0$, but in this limit the correction term diverges. Hence, the ratio $S/U$ remains always finite. Of course, the low temperature expansion applies only if the correction term is significantly smaller than the dominant term in $U$. Still, it is straightforward to show that in the low temperature regime,
  $S/U < 4/T$.

When the box is adiabatically lowered, $T \sim \epsilon$, and the correction term in $U$ also decreases with $\epsilon^{1/3}$. Hence, for a sufficiently dense gas ($N_l \rightarrow \infty$) the first term always dominates. Hence, the ratio $S/U$ diverges as $\epsilon \rightarrow 0$, and we find that the Bekenstein bound is violated for
 \vspace*{-\baselineskip}
\be
x_1 < 0.02 \frac{A^{1/2}S^{3/2}}{H_x^{3/2} N_l^2} (GM)^{1/2}.
\ee
In principle, the violation can take place at macroscopic distance from the horizon if $M$ is taken sufficiently large, but, expectedly, it is much weaker than the corresponding violations at high temperature.



\end{appendix}


\begin{thebibliography}{}

\bibitem{Bek1} J. D. Bekenstein, {\em Black Holes and Entropy},  Phys. Rev. D7, 2333 (1973). %


\bibitem{Hawk} S. W. Hawking, {\em Particle Creation by Black Holes}, Comm. Math. Phys. 43, 19 (1975). %

\bibitem{Bek2} J. D. Bekenstein,  {\em Generalized Second Law of Thermodynamics in Black-Hole Physics},  Phys.
Rev. D9, 3292 (1974). %

\bibitem{Bek3} J. D. Bekenstein, {\em Universal Upper Bound on the Entropy-to-Energy Ratio for Bounded Systems}, Phys. Rev. D23, 287  (1981).

\bibitem{Bek4} J.D. Bekenstein and M. Schiffer, {\em Quantum Limitations on the Storage and Transmission of Information}, Int. J. Mod. Phys. C 1, 355 (1990).

\bibitem{Bek5} J.D. Bekenstein, {\em Do We Understand Black-Hole Entropy?}, in "Proceedings of the VII Marcel Grossmann
Meeting on General Relativity", edited by R. T. Jantzen and G.
M. Keiser (World Scientific, Singapore, 1996).

\bibitem{Bousso} R. Bousso, {\em A Covariant Entropy Conjecture}, JHEP 07, 004 (1999).

\bibitem{AnSav14} N. Savvidou and C. Anastopoulos, {\em The Thermodynamics of Self-gravitating Systems in Equilibrium is Holographic}, Class. Quant. Grav. 31, 055003 (2014).



 \bibitem{UnWa1} W. G. Unruh and R. M. Wald, {\em Acceleration Radiation and the Generalized Second Law of Thermodynamics}, Phys. Rev. D 25, 942 (1982).


\bibitem{Bek8} J. D. Bekenstein, {\em Entropy Bounds and the Second Law for Black Holes}, Phys. Rev. D27, 2262 (1983).




\bibitem{WaldBH} R. M. Wald, {\em Black Holes and Thermodynamics},arXiv:gr-qc/9702022.

\bibitem{DRAW} T. Dauxois, S.Ruffo, E. Arimondo and Martin Wilkens (editors), {\em Dynamics and Thermodynamics of Systems
with Long-Range Interactions} (Springer, Berlin 2002).


\bibitem{Pad1} S. Kolekar and T. Padmanabhan, {\em Ideal gas in a Strong Gravitational Field: Area Dependence of Entropy },  Phys. Rev. D 83, 064034 (2011).

\bibitem{Pad2} S. Bhattacharya, S. Chakraborty, and T. Padmanabhan, {\em Entropy of a Box of Gas in an External Gravitational Field Revisited }, Phys. Rev. D 96, 084030  (2017).

\bibitem{martinez} D. J. Louis-Martinez, {\em Classical Relativistic Ideal Gas in Thermodynamic Equilibrium in a Uniformly Accelerated Reference Frame}, Class. Quantum Grav. 28, 035004 (2011).

\bibitem{thooft} G. 't Hooft, {\em On the Quantum Structure of a Black Hole}, Nucl. Phys. 256B, 727 (1985).


\bibitem{AnSav16} C. Anastopoulos and N. Savvidou, {\em The Thermodynamics of a Black Hole in Equilibrium Implies the Breakdown of Einstein Equations on a Macroscopic Near-Horizon Shell}, JHEP, 2016-144.




\bibitem{Davies}  P. C. W. Davies, {\em Thermodynamics of Black Holes}, Rep. Prog. Phys. 41, 1313 (1978).

\bibitem{Hawk76} S W. Hawking, {\em  Black Holes and Thermodynamics}, Phys. Rev. D 13, 191 (1976).%

\bibitem{Page2} D. N. Page, {\em Black hole formation in a  box},  Gen. Rel. Grav. 13, 1117 (1981).%

\bibitem{York}J.W. York Jr., {\em Black hole Thermodynamics and the Euclidean Einstein Action}, Phys. Rev.D33,  2092 (1986).

\bibitem{LiLi92} L-X Li and L. Liu, {\em Properties of Radiation near the Black-Hole Horizon and the Second Law of Thermodynamics}, Phys. Rev. D 46, 3296 (1992).

\bibitem{Unruhreview}   L. C. B. Crispino, A. Higuchi and G. E. A. Matsas, {\em The Unruh Effect and its Applications}, Rev. Mod. Phys. 80, 778 (2008).



\bibitem{Huang} K. Huang, {\em Statistical Mechanics} (Wiley, 1991).


\bibitem{Visser} M. Visser, {\em Dirty Blackholes: Thermodynamics and Horizon Structure}, Phys. Rev. D46, 2445
 (1992).

\bibitem{Callen} H. B. Callen,  {\em Thermodynamics and an Introduction to Thermostatistics} (John Wiley, New York, 1985).



\bibitem{thooft2} G. 't Hooft, {\em The Scattering Matrix Approach for the Quantum Black Hole, an Overview},  	Int. J. Mod. Phys. A11, 4623 (1996).



\bibitem{KoAn21} D. Kotopoulis and C. Anastopoulos, {\em Thermodynamics and Phase Transitions of Black Holes in Contact with a Gravitating Heat Bath}, Class. Quantum Grav. 38,  195026 (2021).

 \bibitem{AnSav21} C. Anastopoulos and N. Savvidou, {\em Classification Theorem and Properties of Singular Solutions to the Tolman-Oppenheimer-Volkoff Equation}, Class. Quant. Grav. (2021).%

\bibitem{nohv1} M. S. Volkov and D. V. Gal’tsov, {\em Gravitating Non-Abelian Solitons and Black Holes with Yang-Mills Fields},  	Phys. Rep. 319, 1 (1999).

\bibitem{nohv2}P. T. Chrusciel, J. L. Costa and M. Heusler, {\em Stationary Black Holes: Uniqueness and Beyond}, Living Rev. Relativity, 15, 7 (2012).

\bibitem{Israel} S. Mukohyama and W. Israel, {\em Black Holes, Brick Walls, and the Boulware State}, Phys. Rev. D58, 104005 (1998).

\bibitem{Bek6} J. D. Bekenstein, {\em How Does the Entropy/Information Bound Work ?}, Found. Phys. 35, 1805 (2005).

\bibitem{Bousso2} R. Bousso, {\em Bound States and the Bekenstein Bound}, JHEP 0402, 025 (2004).


\bibitem{LWXY} D. Li, B. Wu, Z-M Xu, and W-L Yang, {\em A Shell of Bosons in Spherically Symmetric Spacetimes}, Phys. Lett. B820, 136588 (2021).

\bibitem{fafa} 
F. Scardigli and R. Casadi, {\em Gravitational Tests of the Generalized Uncertainty Principle}, Eur. Phys. J. C  75, 425 (2015).

\bibitem{SMG}M. Soffel, B. M\"uller, and W. Greiner, {\em Dirac Particles in Rindler Space}, Phys. Rev. D22, 1935 (1980).

\bibitem{Oritti} D. Oriti, {\em The Spinor Field in Rindler Spacetime: an Analysis of the Unruh Effect}, Nuovo Cim. B115, 1005 (2000).


\end{thebibliography}
\end{document}